\pdfoutput=1

\documentclass[aps,prd,amsmath,notitlepage,superscriptaddress,%
               nobibnotes,nofootinbib,preprintnumbers,%
               onecolumn,12pt,tightenlines]{revtex4}

\usepackage{graphicx}

\usepackage{verbatim}

\graphicspath{{figures/}}

\usepackage[normalem]{ulem}
\usepackage[usenames]{color}
\definecolor{DarkRed}{rgb}{0.5,0.0,0.0}
\definecolor{DarkGreen}{rgb}{0.0,0.5,0.0}
\definecolor{DarkBlue}{rgb}{0.0,0.0,0.5}
\definecolor{Magenta}{rgb}{1.0,0.0,1.0}
\definecolor{DarkMagenta}{rgb}{0.5,0.0,0.5}
\definecolor{Orange}{rgb}{1.0,0.5,0.0}
\definecolor{DarkOrange}{rgb}{0.8,0.3,0.0}
\definecolor{DarkCyan}{cmyk}{1.0,0.0,0.0,0.5}
\definecolor{Brown}{cmyk}{0.0,0.8,1,0.6}

\usepackage{soul}

\newcommand{\modelLabels}{\bf}
\newcommand{\sigColor}{\textcolor{red}}

\newcommand{\refeqn}[2][eqn:]{Eqn.~(\ref{#1#2})}

\newcommand{\reftab}[2][tab:]{Table~\ref{#1#2}}
\newcommand{\Reftab}[2][tab:]{Table~\ref{#1#2}}
\newcommand{\reffig}[2][fig:]{Figure~\ref{#1#2}}

\newcommand{\refapp}[2][sec:]{the Appendix}
\newcommand{\Refapp}[2][sec:]{The Appendix}

\newcommand{\dof}{\textit{dof}}

\newcommand{\dRdE}{\ensuremath{\frac{dR}{dE}}}

\usepackage{hyperref}

\begin{document}


\title{Examining the time dependence of DAMA's modulation amplitude}

\author{Chris Kelso}
\email[]{ckelso@unf.edu}
\affiliation{
  Department of Physics,
  University of North Florida,
  Jacksonville, FL 32224
  }

\author{Christopher Savage}
\email[]{chris@savage.name}
\affiliation{
  Nordita,
  KTH Royal Institute of Technology and Stockholm University,
  SE-106 91 Stockholm, Sweden}
\affiliation{
  Department of Physics \& Astronomy,
  University of Utah,
  Salt Lake City, UT 84112}

\author{Pearl Sandick}
\email[]{sandick@physics.utah.edu}
\affiliation{
  Department of Physics \& Astronomy,
  University of Utah,
  Salt Lake City, UT 84112}

\author{Katherine Freese}
\email[]{ktfreese@nordita.org}
\affiliation{
  Nordita,
  KTH Royal Institute of Technology and Stockholm University,
  SE-106 91 Stockholm, Sweden}
\affiliation{
  Michigan Center for Theoretical Physics,
  Department of Physics,
  University of Michigan,
  Ann Arbor, MI 48109}

\author{Paolo Gondolo}
\email[]{paolo.gondolo@utah.edu}
\affiliation{
  Department of Physics \& Astronomy,
  University of Utah,
  Salt Lake City, UT 84112}

\date{\today}
 



\begin{abstract} 

If dark matter is composed of weakly interacting particles, Earth's orbital motion may induce a small annual variation in the rate at which these particles interact in a terrestrial detector.  The DAMA collaboration has identified at a 9.3$\sigma$ confidence level such an annual modulation in their event rate over two detector iterations, DAMA/NaI and DAMA/LIBRA, each with $\sim7$ years of observations.  We statistically examine the time dependence of the modulation amplitudes, which "by eye" appear to be decreasing with time in certain energy ranges.  We perform a chi-squared goodness of fit test of the average modulation amplitudes measured\ by the two detector iterations which rejects the hypothesis of a consistent modulation amplitude at greater than 80\%, 96\%, and 99.6\% for the  2--4~keVee, 2--5~keVee and 2--6~keVee energy ranges, respectively.  We also find that among the 14 annual cycles there are three $\gtrsim 3\sigma$ departures from the average in the 5-6~keVee energy range. In addition, we examined several phenomenological models for the time dependence of the modulation amplitude.  Using a maximum likelihood test, we find that descriptions of the modulation amplitude as decreasing with time are preferred over a constant modulation amplitude at anywhere between 1$\sigma$ and 3$\sigma$, depending on the phenomenological model for the time dependence and the signal energy range considered.
A time dependent modulation amplitude is not expected for a dark matter signal, at least for dark matter halo morphologies consistent with the DAMA signal. New data from DAMA/LIBRA--phase2 will certainly aid in determining whether any apparent time dependence is a real effect or a statistical fluctuation.

\end{abstract} 

\maketitle



\section{Introduction}
\label{sec:Intro}
The nature of dark matter is one of the most compelling mysteries in both cosmology and particle physics. One of the foremost possibilities is that dark matter is a new fundamental particle that is yet to be discovered.  There is currently an extensive experimental effort to try to detect the particle nature of dark matter through its interactions with Standard Model particles \cite{Drees:2012ji,Beringer:1900zz}.  Among the leading dark matter candidates are Weakly Interacting Massive Particles (WIMPs), a generic class of particles that includes, for example, the supersymmetric neutralino. These particles interact gravitationally and through the weak force, and their expected masses range from $\mathcal{O}(1)$~GeV to $\mathcal{O}(10)$~TeV.  More than thirty years ago, Refs.~\cite{Drukier:1983gj, Goodman:1984dc} proposed a mechanism for detecting weakly interacting particles, including WIMPs, via coherent elastic scattering with nuclei.  Soon after,  the dark matter 
detection rates in the context of a Galactic Halo of WIMPs were computed for the first time, and
it was proposed that they could be differentiated from background by looking for an annual modulation of the signal~\cite{Drukier:1986tm}. 

The DAMA collaboration has constructed one of these  ``direct detection'' experiments to search for the annual modulation of nuclear recoils due to dark matter scattering within NaI scintillation detectors.  The original detectors were deployed in 1995 and consisted of 100~kg of NaI  \cite{Bernabei:1998rv}.  The experiment went through an upgrade in 2002 that included increasing the detector mass to 250~kg and improving the photo-multiplier tubes that detect the scintillation light from the nuclear recoils.  Through the seven years that the original DAMA/NaI experiment ran, the total exposure was $1.08\cdot 10^5~$kg-days \cite{Bernabei:2000qi,Bernabei:2003za}.  The upgraded DAMA/LIBRA experiment collected a larger exposure of $3.80\cdot 10^5~$kg-days over the seven years it took data \cite{Bernabei:2008yi,Bernabei:2010mq,Bernabei:2013xsa}.  The DAMA collaboration has identified at a 9.3$\sigma$ confidence level  an annual modulation in their event rate that is consistent with a dark matter signal.  Thus far, no plausible alternative explanation for the modulation signal observed in the DAMA detectors has been universally accepted~\cite{Bernabei:2012wp}.  Meanwhile, there are three upcoming NaI experiments that will test the DAMA modulation results: SABRE~\cite{Shields:2015wka}, which will consist of two identical experiments located in the northern and southern hemispheres; ANAIS-112~\cite{Amare:2016fmp, Coarasa:2017aol}, which, as of April 2017, is commissioning at the Canfranc Underground Laboratory; and COSINE-100~\cite{cosine}, a joint effort between the DM-Ice~\cite{deSouza:2016fxg} and KIMS~\cite{Kim:2012rza} collaborations, which has been running at Yangyang Underground Laboratory since September 2016.  Furthermore, DAMA/LIBRA has been running in an upgraded phase 2 configuration since January 2011~\cite{Bernabei:2013qx}, and may also shed light on past DAMA modulation results~\cite{Kelso:2013gda}.

In this work we examine the amplitude of the modulation signal observed by the DAMA collaboration, focusing on the possibility of any time dependence of the amplitude over the 14 years of operation of the DAMA/NaI and DAMA/LIBRA experiments.  We find that descriptions of the modulation amplitude as decreasing with time are preferred over a constant modulation amplitude at anywhere between 1$\sigma$ and 3$\sigma$, depending on the phenomenological model for the time dependence and the signal energy range considered.

In Section \ref{sec:Modulation} we remind the reader of the expected form for the modulation signal.  In Section \ref{sec:Analysis} we present our analysis of the modulation amplitudes and their potential time dependence.  We discuss the results and draw conclusions in Section \ref{sec:Conclusions}.

\section{Annual modulation}
\label{sec:Modulation}

If dark matter is composed of a new, as-yet-unknown particle with weak
interactions with ordinary matter, then dark matter particles in the
galactic halo will pass through the Earth and occasionally scatter on nuclei, causing a nucleus to recoil.
Dark matter direct detection experiments aim to observe these rare,
low-energy nuclear recoils induced by collisions with a dark matter
particle~\cite{Goodman:1984dc}.
The diffuse dark matter halo has little bulk rotation so, from the
perspective of the Earth as it orbits about the center of the galaxy
(along with the rest of the galactic disk) and through the halo,
there is a wind of dark matter approaching from the direction of the
disk rotation.  The form of the expected signal (measured in counts per mass per time per energy) in a direct detection experiment is often written as 

\begin{equation} \label{eqn:Sm}
  \dRdE(E,t) = S_0(E) + S_m(E) \cos[\omega(t-t_0)] + \ldots,
\end{equation}
where $\omega\equiv2\pi$/year, $S_0$ is the time-averaged differential
recoil rate, $S_m$ is the annual modulation amplitude, and higher-order
terms in the expansion are suppressed.  The phase of the expansion $t_0$
is approximately the time of year at which the Earth is moving fastest
into the dark matter wind, around the beginning of June for the
Standard Halo Model (SHM), a common first-order approximation to the
dark matter halo \cite{Drukier:1986tm,Freese:1987wu}.  For a recent review of status of the search for the annual modulation signal from dark matter including alternatives to the SHM, see Ref.~\cite{Freese:2012xd}.

\section{Analysis}
\label{sec:Analysis}

\subsection{DAMA/NaI vs.\ DAMA/LIBRA}
\label{sec:MultiYear}

\begin{table*}
  \begin{center}
  \addtolength{\tabcolsep}{0.5em}
  \begin{tabular}{llccc}
    \hline \hline 
    & & DAMA/NaI & DAMA/LIBRA & Combined \\
    \hline 
    \multicolumn{5}{l}{Exposure [kg-day]} \\
      & & $1.08 \times\!10^{5}$ & $3.80 \times\!10^{5}$ & $4.87 \times\!10^{5}$ \\
    \multicolumn{5}{l}{Average efficiency} \\
      & 2--4 keVee & 0.54 & 0.61 & 0.59 \\
      & 2--5 keVee & 0.63 & 0.65 & 0.65 \\
      & 2--6 keVee & 0.70 & 0.69 & 0.69 \\
      & 4--6 keVee$^\dagger$ & 0.85 & 0.78 & 0.80 \\
      & 4--5 keVee$^\dagger$ & 0.80 & 0.74 & 0.75 \\
      & 5--6 keVee$^\dagger$ & 0.91 & 0.82 & 0.84 \\
    \multicolumn{5}{l}{Modulation amplitude, fixed period \& phase [/kg/day/keVee]} \\
      & 2--4 keVee & 0.0233$\,\pm\,$0.0047 & 0.0167$\,\pm\,$0.0022 & 0.0179$\,\pm\,$0.0020 \\
      & 2--5 keVee & 0.0210$\,\pm\,$0.0038 & 0.0122$\,\pm\,$0.0016 & 0.0135$\,\pm\,$0.0015 \\
      & 2--6 keVee & 0.0192$\,\pm\,$0.0031 & 0.0096$\,\pm\,$0.0013 & 0.0110$\,\pm\,$0.0012 \\
      & 4--6 keVee$^\dagger$
                   & 0.0166$\,\pm\,$0.0041 & 0.0041$\,\pm\,$0.0016 & 0.0059$\,\pm\,$0.0015 \\
      & 4--5 keVee$^\dagger$
                   & 0.0179$\,\pm\,$0.0063 & 0.0048$\,\pm\,$0.0022 & 0.0066$\,\pm\,$0.0022 \\
      & 5--6 keVee$^\dagger$
                   & 0.0155$\,\pm\,$0.0054 & 0.0034$\,\pm\,$0.0022 & 0.0052$\,\pm\,$0.0019 \\
    \multicolumn{5}{l}{Modulation amplitude, free period \& phase [/kg/day/keVee]}  \\
      & 2--4 keVee & 0.0252$\,\pm\,$0.0050 & 0.0178$\,\pm\,$0.0022 & 0.0190$\,\pm\,$0.0020 \\
      & 2--5 keVee & 0.0213$\,\pm\,$0.0039 & 0.0127$\,\pm\,$0.0016 & 0.0140$\,\pm\,$0.0015 \\
      & 2--6 keVee & 0.0200$\,\pm\,$0.0032 & 0.0097$\,\pm\,$0.0013 & 0.0112$\,\pm\,$0.0012 \\
      & 4--6 keVee$^\dagger$
                   & 0.0167$\,\pm\,$0.0042 & 0.0034$\,\pm\,$0.0016 & 0.0054$\,\pm\,$0.0015 \\
      & 4--5 keVee$^\dagger$
                   & 0.0161$\,\pm\,$0.0062 & 0.0043$\,\pm\,$0.0022 & 0.0062$\,\pm\,$0.0022 \\
      & 5--6 keVee$^\dagger$
                   & 0.0173$\,\pm\,$0.0056 & 0.0026$\,\pm\,$0.0022 & 0.0047$\,\pm\,$0.0019 \\
    \hline \hline 
  \end{tabular}
  \end{center}
  \caption[DAMA data and fits]{%
    Best-fit modulation amplitudes for the DAMA/NaI, DAMA/LIBRA, and
    combined exposures.
    Energy-interval-averaged efficiencies are calculated from the overall efficiencies in Fig.~17 of 
    Ref.~\cite{Bernabei:1998rv} and Fig.~26 in Ref.~\cite{Bernabei:2008yh}.
    Modulation amplitudes for the 2--4, 2--5, \& 2--6~keVee energy ranges are fits from
    the DAMA collaboration, taken from
    Refs.~\cite{Bernabei:2003za,Bernabei:2008yh,Bernabei:2013xsa}.
    ($\dagger$) Modulation amplitudes for the 4--6, 4--5, \& 5--6~keVee energy ranges
    are estimates derived from the existing analysis intervals, as
    described in the text (caveats apply).
    }
  \label{tab:Sm}
\end{table*}
   
In this subsection, we gather the best fit modulation amplitudes for the DAMA/NaI and DAMA/LIBRA exposures, as well as the combined best fit amplitudes for the entire 14 years of operation for which the modulation amplitudes have been published.  The DAMA collaboration provides the best fit modulation amplitudes for the 2--4, 2--5, and 2--6 keVee energy ranges. However, we would like to explore the signal in more detail, specifically the higher end of the energy range.  

An estimation for the modulation amplitudes in the  4--6, 4--5, and 5--6 keVee energy ranges, which are not provided in the DAMA collaboration publications, can be obtained from the known modulation amplitudes available from the collaboration using the following procedure. For each energy interval, there is a modulation amplitude ($S_m$), as well as an average efficiency for the detector to detect a recoil ($\varepsilon$).  If this interval is split into two ranges, one can write
\begin{equation}
\varepsilon S_m=f_1\varepsilon_1 S_{m_1}+f_2\varepsilon_2 S_{m_2},
\label{eqn:Sm1}
\end{equation}
with the subscripts 1 and 2 referring to the two energy ranges, and $f_{1,2}$ representing the fraction of the energy interval that each of the two ranges covers. For example, $S_m$ and $S_{m_1}$ could be the published 2-5 keVee and 2-4 keVee amplitudes, and $S_{m_2}$ could be estimated using Eqn.~\ref{eqn:Sm1}. This relationship comes from the fact that the DAMA experiment is fundamentally a counting experiment with the number of events in each range proportional to the total exposure.  
Similarly, the uncertainty ($\sigma$) can be expressed as
\begin{equation}
(\varepsilon \sigma)^2=(f_1\varepsilon_1 \sigma_1)^2+(f_2\varepsilon_2 \sigma_2)^2,
\label{eqn:sigma1}
\end{equation}
with $\sigma_{1,2}$ representing the uncertainty in each of the relevant energy ranges.  This procedure assumes the data for the different energy ranges modulate with the same period and phase but statistically independent, i.e. the 2--5~keVee fit is the appropriately weighted average of fits to the 2--4~keVee and 4--5~keVee ranges.  In the case where the phase and period are fixed, this assumption is valid.  When the phase and period are allowed to vary, however, the assumption is not strictly correct. In looking at the fits provided by the collaboration in Tables 3 and 4 of Ref.~\cite{Bernabei:2013xsa}, we see that the largest difference in modulation amplitude between the fixed versus free fits is at the $\sim0.5\sigma$ level. We thus conclude that the assumption that the energy bins modulate with the smae period and phase is approximately correct to quite good accuracy. This caveat should be kept in mind, however, when examining our results for the derived data in the higher energy ranges. 

\Reftab{Sm} presents the results that have been released to date by the DAMA collaboration along with our estimated data for the higher energy ranges.   The best-fit modulation amplitudes are listed under the assumption of a fixed period and phase, as well as allowing for a free period and phase.  The modulation amplitudes displayed for the 2--4, 2--5, and 2--6~keVee energy ranges are the fits performed by the collaboration as found in Refs.~\cite{Bernabei:2003za,Bernabei:2008yh,Bernabei:2013xsa}, while the modulation amplitudes for the 4--6, 4--5, and 5--6 keVee energy ranges, as well as the efficiencies, are calculated as explained above, and we use the data in Refs.~\cite{Bernabei:1998rv,Bernabei:2008yh} to calculate the average efficiencies.

In comparing the amplitudes for DAMA/NaI to DAMA/LIBRA as shown in \reftab{Sm}, we find that the amplitudes decrease in every energy range.  This points towards a possible inconsistency in the results between the two incarnations of the experiment.  To test the hypothesis of a time-varying modulation amplitude, we begin by examining the compatibility of the DAMA/NaI and DAMA/LIBRA fit results of \reftab{Sm}, under the assumption of a common modulation amplitude.  In \reftab{GOF}, we present the minimum $\chi^2$ and corresponding $p$-value for a $\chi^2$ goodness-of-fit, as well as the best-fit common amplitude.  We find the 2--6~keVee energy range is discrepant at the $\sim$$3\sigma$ level for both the fixed and free period and  phase.  Our estimated data indicates that the majority of the cause of the poor fit is 
due to events with energies $\gtrsim4$ keVee, as the 4--6~keVee interval is similarly discrepant.  Note that the best fit modulation amplitudes in \reftab{GOF} match very well the values presented by the collaboration for the combined fits in the third column of \reftab{Sm}, as expected.

\begin{table*}
  \begin{center}
  \addtolength{\tabcolsep}{0.5em}
  \begin{tabular}{llccc}
    \hline \hline 
    & & $\chi^2$/\dof & $p$ & $S_m$ \\
    \hline 
    \multicolumn{5}{l}{Fixed period \& phase} \\
      & 2--4 keVee & 1.61/1 & 0.203 (1.3$\sigma$) & 0.0179$\,\pm\,$0.0020 \\
      & 2--5 keVee & 4.55/1 & 0.033 (2.1$\sigma$) & 0.0135$\,\pm\,$0.0015 \\
      & 2--6 keVee & 8.16/1 & 0.004 (2.9$\sigma$) & 0.0110$\,\pm\,$0.0012 \\
      & 4--6 keVee$^\dagger$
                   & 8.19/1 & 0.004 (2.9$\sigma$) & 0.0057$\,\pm\,$0.0015 \\
      & 4--5 keVee$^\dagger$
                   & 3.87/1 & 0.049 (1.9$\sigma$) & 0.0063$\,\pm\,$0.0021 \\
      & 5--6 keVee$^\dagger$
                   & 4.34/1 & 0.037 (2.1$\sigma$) & 0.0054$\,\pm\,$0.0020 \\
    \multicolumn{5}{l}{Free period \& phase}  \\
      & 2--4 keVee & 1.84/1 & 0.176 (1.4$\sigma$) & 0.0190$\,\pm\,$0.0020 \\
      & 2--5 keVee & 4.16/1 & 0.041 (2.0$\sigma$) & 0.0139$\,\pm\,$0.0015 \\
      & 2--6 keVee & 8.89/1 & 0.003 (3.0$\sigma$) & 0.0112$\,\pm\,$0.0012 \\
      & 4--6 keVee$^\dagger$
                   & 9.01/1 & 0.003 (3.0$\sigma$) & 0.0050$\,\pm\,$0.0015 \\
      & 4--5 keVee$^\dagger$
                   & 3.18/1 & 0.074 (1.8$\sigma$) & 0.0057$\,\pm\,$0.0021 \\
      & 5--6 keVee$^\dagger$
                   & 6.01/1 & 0.014 (2.5$\sigma$) & 0.0045$\,\pm\,$0.0020 \\
    \hline \hline 
  \end{tabular}
  \end{center}
  \caption[DAMA goodness-of-fit]{%
    Compatibility of the DAMA/NaI and DAMA/LIBRA fit results of
    \reftab{Sm}, under the assumption of a common modulation amplitude.
    The minimum $\chi^2$ and corresponding $p$-value for a $\chi^2$
    goodness-of-fit are shown, as well as the best-fit common amplitude.
    ($\dagger$) Data for these energy intervals are derived as discussed
    in the text.
    }
  \label{tab:GOF}
\end{table*}

We also collect the fits for the per-cycle modulation amplitudes as performed by the DAMA collaboration for the free period and phase in \reftab{CycleSm}.   Unfortunately the fits to the data under the assumption of a fixed period and phase have not been released to the public, and we are thus unable to analyze the data in that case.  The mean cycle times and amplitudes are taken from Figure~3 of Ref.~\cite{Bernabei:2013xsa} and exposures are given in Refs.~\cite{Bernabei:2003za,Bernabei:2013xsa}. The mean time for each cycle is relative to January~1, 1995, the first year of DAMA/NaI's operation.

\begin{table*}
  \begin{center}
  \addtolength{\tabcolsep}{0.5em}
  \begin{tabular}{llccr@{$\,\pm\,$}lr@{$\,\pm\,$}lr@{$\,\pm\,$}l}
    \hline \hline 
    & & Exposure  & $\langle t\rangle$
      & \multicolumn{2}{c}{$S_m$ (2--4~keVee)}
      & \multicolumn{2}{c}{$S_m$ (2--5~keVee)}
      & \multicolumn{2}{c}{$S_m$ (2--6~keVee)} \\
    & & [kg-days] & [year]
      & \multicolumn{2}{c}{[/kg/day/keVee]}
      & \multicolumn{2}{c}{[/kg/day/keVee]}
      & \multicolumn{2}{c}{[/kg/day/keVee]} \\
    \hline 
    \multicolumn{10}{l}{DAMA/NaI} \\
    & cycle~1 &  4549 &  1.19  &  0.005 & 0.021  &  0.017 & 0.016  &  0.024 & 0.012 \\
    & cycle~2 & 14962 &  2.23  &  0.022 & 0.012  &  0.019 & 0.009  &  0.023 & 0.008 \\
    & cycle~3 & 22455 &  3.15  &  0.017 & 0.012  &  0.012 & 0.010  &  0.025 & 0.008 \\
    & cycle~4 & 16020 &  4.17  &  0.023 & 0.013  &  0.023 & 0.010  &  0.018 & 0.009 \\
    & cycle~5 & 15911 &  5.10  &  0.038 & 0.013  &  0.025 & 0.010  &  0.017 & 0.009 \\
    & cycle~6 & 16608 &  6.20  &  0.021 & 0.011  &  0.021 & 0.009  &  0.011 & 0.007 \\
    & cycle~7 & 17226 &  7.06  &  0.028 & 0.012  &  0.028 & 0.009  &  0.021 & 0.008 \\
    \multicolumn{10}{l}{DAMA/LIBRA} \\
    & cycle~1 & 51405 &  9.12  &  0.0294 & 0.0064  &  0.0181 & 0.0048  &  0.0098 & 0.0038 \\
    & cycle~2 & 52597 & 10.19  &  0.0194 & 0.0072  &  0.0134 & 0.0053  &  0.0089 & 0.0044 \\
    & cycle~3 & 39445 & 11.19  &  0.0172 & 0.0062  &  0.0173 & 0.0045  &  0.0122 & 0.0038 \\
    & cycle~4 & 49337 & 12.04  &  0.0192 & 0.0061  &  0.0181 & 0.0045  &  0.0125 & 0.0036 \\
    & cycle~5 & 66105 & 13.11  &  0.0105 & 0.0053  &  0.0091 & 0.0041  &  0.0086 & 0.0031 \\
    & cycle~6 & 58768 & 14.27  &  0.0150 & 0.0052  &  0.0070 & 0.0039  &  0.0102 & 0.0030 \\
    & cycle~7 & 62098 & 15.17  &  0.0166 & 0.0050  &  0.0097 & 0.0036  &  0.0092 & 0.0028 \\
    \hline \hline 
  \end{tabular}
  \end{center}
  \caption[Modulation amplitudes]{%
    The per-cycle modulation amplitudes as determined by the DAMA collaboration, allowing
    the period and phase to be freely fit in addition to the amplitude.
    Mean cycle times and amplitudes are taken from Figure~3 of
    Ref.~\cite{Bernabei:2013xsa} and exposures are taken from
    Refs.~\cite{Bernabei:2003za,Bernabei:2013xsa}.
    The mean time for each cycle is given relative to January~1, 1995,
    the first year of DAMA/NaI's operation.
    }
  \label{tab:CycleSm}
\end{table*}

There are many different possible run tests that can be performed on a given set of data to check for consistency. In this context, a run is a group of consecutive identical elements in a two-valued random sequence constructed from the data. A run test checks for runs in the data, in terms of a two-valued characteristic one can assign to each data point. The collaboration has performed run tests to determine whether all data are consistent with the best fit value for the amplitude, and found lower tail probabilities of 41\%, 29\%, and 23\% for the 2--4, 2--5, and 2--6~keVee energy ranges, respectively. We have repeated the DAMA collaboration run tests and reproduced their results (see Table~\ref{tab:runTest}; Above and Below Best Fit). Here, we are interested in the consistency between the two iterations of the DAMA experiment, NaI and LIBRA, and we  perform several other tests of interest as described below.    

For the DAMA data, there are a total of 7 data points for each of the two iterations of the experiment  (for each energy bin).  To check whether the NaI and LIBRA data are consistent with each other, we rank the full 14 measurements from low to high, keeping track of which experiment made the given measurement.  The number of runs is then one plus the number of times in this list the subsequent element changes from one version of the experiment to the other.  As an example, when the amplitude values for the 2--6~keVee energy range are sorted in this way, the list would be
(LIBRA, LIBRA, LIBRA, LIBRA, LIBRA, NaI, LIBRA, LIBRA, NaI, NaI, NaI, NaI, NaI, NaI), 
giving a total of four runs in this list.  In this version of the run test, the null hypothesis would be that the two distributions are equal, i.e. that the DAMA/NaI and DAMA/LIBRA experiments are measuring the same amplitude in a given energy bin.  The alternative hypothesis is that the two distributions are not equal. 

Under the null hypothesis, i.e.~that the two populations of data are drawn from the same distribution, the probability of obtaining a certain number of runs can be calculated using
\begin{equation}
P(R=2k)=2\frac{\left( 
\begin{array}{c}
n_1-1\\
k-1
\end{array}\right)
\left( 
\begin{array}{c}
n_2-1\\
k-1
\end{array}\right)}{\left( 
\begin{array}{c}
n_1+n_2\\
n_1
\end{array}\right)},
\end{equation}
where $R$ is the number of observed runs, $n_1$ is the number of values from one population or experiment, and $n_2$ is the number of values from the other population or experiment, and $k$ is an integer that gives the appropriate number of runs. Similarly, if the number of runs is odd, the probability of obtaining that number of runs can be written as
\begin{equation}
P(R=2k+1)=\frac{\left( 
\begin{array}{c}
n_1-1\\
k
\end{array}\right)
\left( 
\begin{array}{c}
n_2-1\\
k-1
\end{array}\right)+\left( 
\begin{array}{c}
n_2-1\\
k
\end{array}\right)
\left( 
\begin{array}{c}
n_1-1\\
k-1
\end{array}\right)}{\left( 
\begin{array}{c}
n_1+n_2\\
n_1
\end{array}\right)}.
\end{equation}
The lower tail probability can then be computed by summing the probability of two runs (there must be at least two runs) up to the actual number of runs observed.  To be concrete, there are four runs in the case of the 2--6~keVee energy bin, so the probability would then be summed for obtaining, two, three, and four runs.  The results for this run test applied to the data in \reftab{CycleSm} are collected in \reftab{runTest} under the heading of Ranked by Experiment.  We find that the null hypothesis for the 2--6~keVee energy range would be rejected at the 2$\sigma$ level, while the 2-4~keVee and 2-5~keVee energy ranges have $p$-values of 30\% and 21\%, respectively.

Other possible run tests focus on fluctuations above and below a reference point, i.e. the median, mean, best fit point, etc.  In this case, the measurements are put in an array in the order in which they were measured.  Then a new list is formed where at each element, a '+' sign or '-' is placed depending on if the measurement is above or below the reference point.  The number of runs are then counted in this list and the probabilities calculated exactly as in the previous case with $n_1$ and $n_2$ now representing the number of points above and below the reference value.  The null hypothesis for this run test is that the data points are randomly fluctuating about the reference value.  A run test of this type with reference to the best fit point is employed by the DAMA collaboration, and we reproduce their $p$-values.  The results for these run tests are presented in \reftab{runTest}.  

\begin{table*}[ht!]
  \begin{center}
  \addtolength{\tabcolsep}{0.5em}
  \begin{tabular}{lccc}
    \hline \hline 
    & & Number of Runs & $p$  \\
    \hline 
		\multicolumn{4}{l}{{\modelLabels Ranked by Experiment} } \\
      & 2--4 keVee
        & 7 & 0.383 (0.30$\sigma$)\\
      & 2--5 keVee
        & 6 & 0.208 (0.81$\sigma$)\\
      & 2--6 keVee
        & 4 & 0.0251 (2.0$\sigma$)\\
				
     \multicolumn{4}{l}{{\modelLabels Above and Below Best Fit} }\\
      & 2--4 keVee
        & 7 & 0.413 (0.22$\sigma$)\\
      & 2--5 keVee
        & 6 & 0.287 (0.56$\sigma$)\\
      & 2--6 keVee
        & 6 & 0.226 (0.75$\sigma$) \\

    \multicolumn{4}{l}{{\modelLabels Above and Below Median} }\\
      & 2--4 keVee
        & 5 & 0.0775 (1.4$\sigma$) \\
      & 2--5 keVee
        & 7 & 0.383 (0.30$\sigma$) \\
      & 2--6 keVee
        & 6 & 0.209 (0.81$\sigma$)\\   
   					
    \multicolumn{4}{l}{{\modelLabels KS Test} }\\
      & 2--4 keVee
        & -- & 0.21 (0.80$\sigma$)\\
      & 2--5 keVee
        & -- & 0.053 (1.6$\sigma$)\\
      & 2--6 keVee
        & -- & 0.0082 (2.4$\sigma$) \\

    \hline \hline 
  \end{tabular}
  \end{center}
  \caption[Run Tests]{%
    Results for the run test as presented by the collaboration and for the tests we have performed as described in the text.  The corresponding $p$-value is for a hypothesis test where the null hypothesis is taken to be the case where the two iterations of the experiment are drawing measurements from the same distribution. 

    }
  \label{tab:runTest}
\end{table*}

We also utilized the Kolmogorov-Smirnov (KS) test for the two data sets.  The KS test is another non-parametric test in which the null hypothesis is that two data sets are drawn from the same distribution.  As can be seen in \reftab{runTest}, we find that the null hypothesis has $p$-values of 21\% and 5\% in the 2-4~keVee and 2-5~keVee energy ranges, respectively, and that it is rejected at the $2.4\sigma$ level in the 2--6~keVee energy range.  One possible explanation for these results, which will be explored in the remainder of this paper, is that the modulation amplitude is changing (seemingly decaying) with time. 

\subsection{Annual cycles}
\label{sec:Cycles}

Given the apparent decrease in the modulation amplitude from DAMA/NaI to DAMA/LIBRA, it is interesting to investigate in detail the time dependence of the modulation amplitude.
Here, we postulate several phenomenological models for the time dependence of the modulation amplitude, and perform likelihood fits to investigate whether or not any model is favored by the data.

\begin{figure}[ht!]
\includegraphics[keepaspectratio,width=\columnwidth]{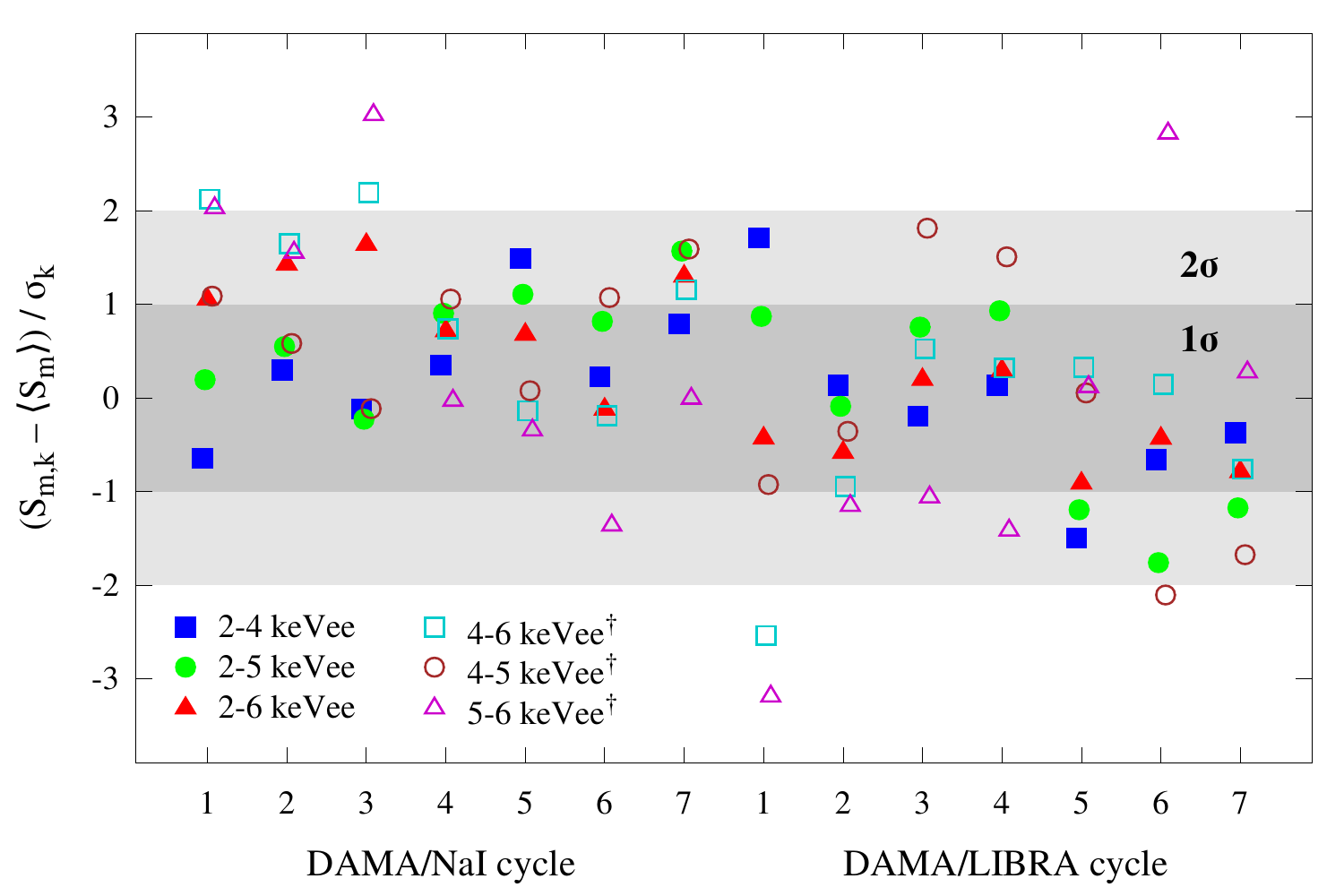}%
\caption{The deviations of the modulation amplitude in each annual cycle for a given energy range compared to the average in that range.  $S_{m,k}$ and $\sigma_k$ are the amplitude and uncertainty in the $k$th annual cycle, respectively, and $\left\langle S_m\right\rangle$ is the average amplitude in the particular energy range for  the entire 14 year period.}%
\label{fig:Deviations}%
\end{figure}

A visual representation of the deviations of the amplitude in each annual cycle from the average is shown in \reffig{Deviations}.  For each energy interval considered, we plot the deviation in each annual cycle as $\frac{S_{m,k}-\left\langle S_m\right\rangle}{\sigma_k}$ 
where $S_{m,k}$ and $\sigma_k$ are the amplitude and uncertainty in the $k$th annual cycle, respectively, and $\left\langle S_m\right\rangle$ is the uncertainty-weighted average amplitude in the particular energy range over the entire 14 year period. We observe that all the published data (in the 2-4~keVee, 2-5~keVee, and 2-6~keVee energy ranges) fall within $2\sigma$ of the average. A Shapiro-Wilk test of normality of the deviations gives $p$-values of 73\%, 10\%, and 58\% for these published energy ranges, respectively. Low $p$-values point to a non-normal distribution of the published deviations.  The most striking feature in the plot is the fact that among the 14 annual cycles, there are three $>2.8\sigma$ departures from the average in the 5-6~keVee energy range.  If the data in this range were gaussian distributed, then the probability of this occurring is $2\cdot 10^{-5}\,(4.2\sigma)$. This seemingly unlikely situation	is an indication that there may be a problem with the data in this energy range.	

To describe a potential time dependence of the modulation amplitude, we have performed likelihood tests for the following models: a constant modulation amplitude, separate constants for NaI and LIBRA, linear in time, exponential in time, and a broken exponential (a common exponential scale, but independent normalizations for NaI and LIBRA). We estimate the per-cycle modulation amplitudes for the 4--6, 4--5, and 5--6~keVee energy ranges using \refeqn{Sm1} and \refeqn{sigma1} and these values are shown in \reftab{CycleSmDerived}. Fits for these models to the per-cycle DAMA modulation amplitude data from \reftab{CycleSm} and \reftab{CycleSmDerived} are presented in \reftab{CyclesGOF}.  The minimum $\chi^2$ and corresponding $p$-value for a $\chi^2$ goodness-of-fit are shown.  Best-fit parameters are found in the final column and parameters not shown are profiled over.  We have highlighted in red any hypotheses that are discrepant at $\geq2.5\sigma$. 

Based on the goodness of fit test, we find that all the models we investigated (including the single constant amplitude) provide reasonable fits to each of the energy ranges, with the exception of the 5--6~keVee range.  This range is excluded at nearly $4\sigma$ by the goodness of fit test for every model we studied.  This is the quantitative conclusion regarding the odd behavior in the 5--6 keVee range observed in \reffig{Deviations}.  

Finally, we compared models (hypothesis tests) against a single, constant modulation amplitude taken as the null hypothesis.  The results are displayed in \reftab{CyclesGOF} with the improvement in fit given by the $\Delta\chi^2$ along with the corresponding $p$-value.  The improvements in fit for the 2--4~keVee energy range are marginal ($1.0\sigma$ to $1.6\sigma$), indicating consistency in this energy range.  We determine, however, that all time-varying models we investigated are preferred over a constant amplitude in the 2--6~keVee energy range at levels from $2.3\sigma$ to $2.7\sigma$.  As the 5--6~keVee data may be somewhat suspect, we find that the alternative models are still preferred over a constant amplitude in the 2-5~keVee range at the level of $2.0\sigma$ to $2.7\sigma$.  Although the data clearly indicate that some form of time dependence is preferred, none of the alternative models is clearly better than the others.  Unfortunately, the data is not capable of distinguishing between the different functional forms proposed here to model the time dependence of the modulation amplitude.

\begin{table*}
  \begin{center}
  \addtolength{\tabcolsep}{0.5em}
  \begin{tabular}{llr@{$\,\pm\,$}lr@{$\,\pm\,$}lr@{$\,\pm\,$}l}
    \hline \hline 
    & & \multicolumn{2}{c}{$S_m$ (4--6~keVee)}
      & \multicolumn{2}{c}{$S_m$ (4--5~keVee)}
      & \multicolumn{2}{c}{$S_m$ (5--6~keVee)} \\
    & & \multicolumn{2}{c}{[/kg/day/keVee]}
      & \multicolumn{2}{c}{[/kg/day/keVee]}
      & \multicolumn{2}{c}{[/kg/day/keVee]} \\
    \hline 
    \multicolumn{8}{l}{DAMA/NaI} \\
    & cycle~1 &  0.036 & 0.014  &  0.033 & 0.025  &  0.039 & 0.016 \\
    & cycle~2 &  0.024 & 0.011  &  0.015 & 0.015  &  0.032 & 0.016 \\
    & cycle~3 &  0.030 & 0.011  &  0.005 & 0.016  &  0.052 & 0.015 \\
    & cycle~4 &  0.014 & 0.011  &  0.023 & 0.016  &  0.007 & 0.016 \\
    & cycle~5 &  0.004 & 0.011  &  0.008 & 0.015  &  0.001 & 0.016 \\
    & cycle~6 &  0.004 & 0.009  &  0.021 & 0.014  & -0.011 & 0.013 \\
    & cycle~7 &  0.017 & 0.010  &  0.028 & 0.014  &  0.007 & 0.013 \\
    \multicolumn{8}{l}{DAMA/LIBRA} \\
    & cycle~1 & -0.0053 & 0.0044  & -0.0003 & 0.0073  & -0.0099 & 0.0053 \\
    & cycle~2 &  0.0008 & 0.0054  &  0.0037 & 0.0076  & -0.0019 & 0.0077 \\
    & cycle~3 &  0.0083 & 0.0046  &  0.0176 & 0.0062  & -0.0001 & 0.0067 \\
    & cycle~4 &  0.0073 & 0.0043  &  0.0163 & 0.0066  & -0.0009 & 0.0056 \\
    & cycle~5 &  0.0071 & 0.0037  &  0.0068 & 0.0062  &  0.0075 & 0.0043 \\
    & cycle~6 &  0.0064 & 0.0034  & -0.0060 & 0.0059  &  0.0176 & 0.0038 \\
    & cycle~7 &  0.0035 & 0.0031  & -0.0016 & 0.0048  &  0.0081 & 0.0041 \\
    \hline \hline 
  \end{tabular}
  \end{center}
  \caption[Modulation amplitudes (derived)]{%
    The per-cycle modulation amplitudes for the 4--6, 4--5,
    and 5--6~keVee energy ranges, derived from the data in \reftab{CycleSm} as described
    in the text.  Exposures and mean times for each cycle are the same
    as in \reftab{CycleSm}.
    }
  \label{tab:CycleSmDerived}
\end{table*}

\begin{table*}
  \begin{center}
  \addtolength{\tabcolsep}{0.5em}
  \begin{tabular}{llccccc}
    \hline \hline 
    & & \multicolumn{2}{c}{goodness-of-fit\hspace*{2em}}
      & \multicolumn{2}{c}{vs.\ null hypothesis\hspace*{2em}}
      & best-fit parameters \\
    & & $\chi^2$/\dof & $p$ & $\Delta\chi^2$/\dof & $p$ & \\
    \hline 
    \multicolumn{6}{l}{{\modelLabels Constant (null hypothesis):} $S_m(t) = A$}
    & \multicolumn{1}{c}{$A$ [dru]} \\
      & 2--4 keVee
        & 9.36/13 & 0.746 (0.3$\sigma$)
        & --      & --
        & 0.0184 $\pm$ 0.0020 \\
      & 2--5 keVee
        & 13.7/13 & 0.396 (0.8$\sigma$)
        & --      & --
        & 0.0139 $\pm$ 0.0015 \\
      & 2--6 keVee
        & 10.9/13 & 0.618 (0.5$\sigma$)
        & --      & --
        & 0.0114 $\pm$ 0.0012 \\
      & 4--6 keVee$^\dagger$
        & 22.4/13 & 0.049 (2.0$\sigma$)
        & --      & --
        & 0.0059 $\pm$ 0.0014 \\
      & 4--5 keVee$^\dagger$
        & 20.1/13 & 0.092 (1.7$\sigma$)
        & --      & --
        & 0.0064 $\pm$ 0.0022 \\
      & 5--6 keVee$^\dagger$
        & 40.4/13 & \sigColor {0.0001 (3.8$\sigma$)}
        & --      & --
        & 0.0069 $\pm$ 0.0018 \\
    \multicolumn{6}{l}{{\modelLabels Broken constant:} $S_m(t) = A_N\ \text{(NaI)},\ A_L\ \text{(LIBRA)}$}
    & \multicolumn{1}{c}{$A_N,A_L$ [dru]} \\
      & 2--4 keVee
        & 8.10/12 & 0.777 (0.3$\sigma$)
        & 1.26/1  & 0.262 (1.1$\sigma$)
        & 0.0234, 0.0174 \\
      & 2--5 keVee
        & 9.19/12 & 0.686 (0.4$\sigma$)
        & 4.49/1  & 0.034 (2.1$\sigma$)
        & 0.0211, 0.0125 \\
      & 2--6 keVee
        & 3.62/12 & 0.989 (0.01$\sigma$)
        & 7.29/1  & \sigColor{0.007 (2.7$\sigma$)}
        & 0.0192, 0.0101 \\
      & 4--6 keVee$^\dagger$
        & 14.2/12 & 0.289 (1.1$\sigma$)
        & 8.25/1  & \sigColor{0.004 (2.9$\sigma$)}
        & 0.0169, 0.0044 \\
      & 4--5 keVee$^\dagger$
        & 15.4/12 & 0.220 (1.2$\sigma$)
        & 4.73/1  & 0.030 (2.2$\sigma$)
        & 0.0182, 0.0045 \\
      & 5--6 keVee$^\dagger$
        & 37.0/12 & \sigColor{0.0002 (3.7$\sigma$)}
        & 3.39/1  & 0.066 (1.8$\sigma$)
        & 0.0168, 0.0059 \\
    \multicolumn{6}{l}{{\modelLabels Linear:} $S_m(t) = mt+b$}
    & \multicolumn{1}{c}{$m$ [$10^{-3}$ dru/yr]} \\
      & 2--4 keVee
        & 6.85/12 & 0.867 (0.2$\sigma$)
        & 2.50/1  & 0.114 (1.6$\sigma$)
        & -0.87 $\pm$ 0.55 \\
      & 2--5 keVee
        & 6.54/12 & 0.886 (0.1$\sigma$)
        & 7.15/1  & \sigColor{0.008 (2.7$\sigma$)}
        & -1.12 $\pm$ 0.42 \\
      & 2--6 keVee
        & 4.40/12 & 0.975 (0.03$\sigma$)
        & 6.51/1  & \sigColor{0.011 (2.6$\sigma$)}
        & -0.87 $\pm$ 0.34 \\
      & 4--6 keVee$^\dagger$
        & 18.8/12 & 0.092 (1.7$\sigma$)
        & 3.58/1  & 0.058 (1.9$\sigma$)
        & -0.80 $\pm$ 0.42 \\
      & 4--5 keVee$^\dagger$
        & 13.5/12 & 0.335 (1.0$\sigma$)
        & 6.65/1  & \sigColor{0.010 (2.6$\sigma$)}
        & -1.63 $\pm$ 0.63 \\
      & 5--6 keVee$^\dagger$
        & 40.3/12 & \sigColor{0.00006 (4.0$\sigma$)}
        & 0.11/1  & 0.736 (0.3$\sigma$)
        &  0.18 $\pm$ 0.55 \\
    \multicolumn{6}{l}{{\modelLabels Exponential:} $S_m(t) = Ae^{-\beta t}$}
    & \multicolumn{1}{c}{$\beta$ [/yr]} \\
      & 2--4 keVee
        & 7.14/12 & 0.849 (0.2$\sigma$)
        & 2.22/1  & 0.136 (1.5$\sigma$)
        & 0.039 $\pm$ 0.025 \\
      & 2--5 keVee
        & 7.59/12 & 0.816 (0.2$\sigma$)
        & 6.10/1  & \sigColor{0.014 (2.5$\sigma$)}
        & 0.061 $\pm$ 0.023 \\
      & 2--6 keVee
        & 3.64/12 & 0.989 (0.01$\sigma$)
        & 7.27/1  & \sigColor{0.007 (2.7$\sigma$)}
        & 0.072 $\pm$ 0.024 \\
      & 4--6 keVee$^\dagger$
        & 14.9/12 & 0.245 (1.2$\sigma$)
        & 7.48/1  & \sigColor{0.006 (2.7$\sigma$)}
        &  0.18 $\pm$ 0.07  \\
      & 4--5 keVee$^\dagger$
        & 15.3/12 & 0.223 (1.2$\sigma$)
        & 4.79/1  & 0.029 (2.2$\sigma$)
        &  0.14 $\pm$ 0.06  \\
      & 5--6 keVee$^\dagger$
        & 35.6/12 & \sigColor{0.0004 (3.6$\sigma$)}
        & 4.79/1  & 0.029 (2.2$\sigma$)
        & -0.39 $\pm$ 0.18  \\
    \multicolumn{6}{l}{{\modelLabels Broken exponential:} $S_m(t) = A_N e^{-\beta t}\ \text{(NaI)},\ A_L e^{-\beta t}\ \text{(LIBRA)}$}
    & \multicolumn{1}{c}{$\beta$ [/yr]} \\
      & 2--4 keVee
        & 6.93/11 & 0.805 (0.2$\sigma$)
        & 2.43/2  & 0.297 (1.0$\sigma$)
        & 0.062 $\pm$ 0.058 \\
      & 2--5 keVee
        & 7.57/11 & 0.751 (0.3$\sigma$)
        & 6.12/2  & 0.047 (2.0$\sigma$)
        & 0.068 $\pm$ 0.058 \\
      & 2--6 keVee
        & 3.03/11 & 0.990 (0.01$\sigma$)
        & 7.88/2  & 0.019 (2.3$\sigma$)
        & 0.038 $\pm$ 0.050 \\
      & 4--6 keVee$^\dagger$
        & 13.6/11 & 0.253 (1.1$\sigma$)
        &  8.78/2  &  \sigColor{0.012 (2.5$\sigma$)}
        &  0.07 $\pm$ 0.11  \\
      & 4--5 keVee$^\dagger$
        & 15.1/11 & 0.178 (1.3$\sigma$)
        & 5.03/2  & 0.081 (1.7$\sigma$)
        &  0.08 $\pm$ 0.14  \\
      & 5--6 keVee$^\dagger$
        & 34.7/11 & \sigColor{0.0003 (3.6$\sigma$)}
        & 5.73/2  & 0.057 (1.9$\sigma$)
        & -0.30 $\pm$ 0.20  \\
    \hline \hline 
  \end{tabular}
  \end{center}
  \caption[DAMA fits]{%
    Fits of the per-cycle DAMA modulation amplitude data of
    \reftab{CycleSm} and \reftab{CycleSmDerived}.
    The minimum $\chi^2$ and corresponding $p$-value for a $\chi^2$
    goodness-of-fit are shown, followed by the $\Delta\chi^2$ and
    corresponding $p$-value for a hypothesis test where the conventional
    constant modulation amplitude is taken as the null hypothesis. 
    Best-fit parameters are shown in the final column; parameters not
    shown are profiled over.   We have highlighted in red any hypotheses that are discrepant at  $\geq2.5\sigma$.
    A differential rate unit (dru) is equal to 1~count/kg/day/keVee.
    ($\dagger$) Data for these energy intervals are derived as discussed
    in the text.
    }
  \label{tab:CyclesGOF}
\end{table*}

\section{Discussion and conclusions}
\label{sec:Conclusions}
In this study, we have analyzed the annual modulation data released by the DAMA collaboration that they collected over 14 annual cycles and performed likelihood analyses for several phenomenological models to explore a possible time dependence in the modulation amplitude.  Our results indicate that all of the models we examined are preferred over a constant amplitude at up to $\sim$$3\sigma$.  Although the data clearly prefer some form of time dependence, they are currently incapable of distinguishing between the different functional forms for the time dependence we investigated.  

Though we have identified the fact that the modulation amplitude is discrepant at the \mbox{2--3$\sigma$} level for each of the two versions of the experiment in all but the 2--4~keVee energy range, we do not propose a physical explanation for the phenomenon.  More data will certainly aid in determining whether this is a real effect or a statistical fluctuation.  The DAMA/LIBRA experiment has undergone a significant upgrade, and has been taking data in the DAMA/LIBRA-phase 2 configuration since January 2011~\cite{Bernabei:2013qx}. In November of 2010, all of the photo-multiplier tubes (PMT's) in the DAMA/LIBRA experiment were replaced by high quantum efficiency PMT's~\cite{1748-0221-7-03-P03009}.  The anticipated lower thresholds ($\sim$1~keVee) will give access to a new signal range that should help to clarify the dark matter interpretation of the DAMA signal, as described in Ref~\cite{Kelso:2013gda}.  

It is also interesting to note that the half life for the exponential decay model in the 2--5~keVee energy range is about 11 years.  If the collaboration releases data again in the near future, both the linear and exponential decay models would predict a noticeable decrease in the modulation amplitude, somewhere in the neighborhood of 50\%.  If this trend continues, this would have serious implications for the dark matter interpretation of the DAMA modulation.  

In this study we have attempted to answer the question of whether the DAMA collaboration data demonstrates any time dependence, and the answer to that question appears to be ``yes,'' at the 2--3$\sigma$ significance level in all but the 2--4~keVee energy range.  A more conclusive answer will of course require additional data, which we look forward to in the near future, from the DAMA collaboration as well as from ANAIS-112, COSINE-100, and SABRE, each of which will test the DAMA modulation results.  Furthermore, given the questions raised here as well as the tension of the DAMA results with results from other dark matter direct detection experiments, we urge the DAMA collaboration and other experimental collaborations to make enough of their data public (i.e.~providing time-stamped events) in order for researchers to repeat their analyses to encourage scrutiny and, with any luck, a consistent explanation for the observed phenomena, dark matter or otherwise.


\acknowledgments
  KF acknowledges support from the Swedish Research Council (Vetenskapsrûadet) through the Oskar Klein Centre (Contract No. 638-2013-8993). KF acknowledges support from DoE grant DE-SC007859 at the University of Michigan. PS is supported in part by NSF Grant No. PHY-1417367. PG was partially supported by NSF Award PHY-1415974.
  CK, PS, and CS thank the Department of Physics \& Astronomy at the
  University of Utah for support. PG, CK, PS, and CS thank Nordita, the Nordic Institute for Theoretical Physics, for its hospitality. This research was supported in part by the Munich Institute for Astro- and Particle Physics (MIAPP) of the DFG cluster of excellence ``Origin and Structure of the Universe.'' We would also like to thank Peter Smith for helpful comments on the draft of this paper.




\bibliography{DAMASm}

\begin{thebibliography}{24}
\expandafter\ifx\csname natexlab\endcsname\relax\def\natexlab#1{#1}\fi
\expandafter\ifx\csname bibnamefont\endcsname\relax
  \def\bibnamefont#1{#1}\fi
\expandafter\ifx\csname bibfnamefont\endcsname\relax
  \def\bibfnamefont#1{#1}\fi
\expandafter\ifx\csname citenamefont\endcsname\relax
  \def\citenamefont#1{#1}\fi
\expandafter\ifx\csname url\endcsname\relax
  \def\url#1{\texttt{#1}}\fi
\expandafter\ifx\csname urlprefix\endcsname\relax\def\urlprefix{URL }\fi
\providecommand{\bibinfo}[2]{#2}
\providecommand{\eprint}[2][]{\url{#2}}

\bibitem[{\citenamefont{Drees and Gerbier}(2012)}]{Drees:2012ji}
\bibinfo{author}{\bibfnamefont{M.}~\bibnamefont{Drees}} \bibnamefont{and}
  \bibinfo{author}{\bibfnamefont{G.}~\bibnamefont{Gerbier}}
  (\bibinfo{year}{2012}), \eprint{1204.2373}.

\bibitem[{\citenamefont{Beringer et~al.}(2012)}]{Beringer:1900zz}
\bibinfo{author}{\bibfnamefont{J.}~\bibnamefont{Beringer}} \bibnamefont{et~al.}
  (\bibinfo{collaboration}{Particle Data Group}), \bibinfo{journal}{Phys.Rev.}
  \textbf{\bibinfo{volume}{D86}}, \bibinfo{pages}{010001}
  (\bibinfo{year}{2012}).

\bibitem[{\citenamefont{Drukier and Stodolsky}(1984)}]{Drukier:1983gj}
\bibinfo{author}{\bibfnamefont{A.}~\bibnamefont{Drukier}} \bibnamefont{and}
  \bibinfo{author}{\bibfnamefont{L.}~\bibnamefont{Stodolsky}},
  \bibinfo{journal}{Phys.Rev.} \textbf{\bibinfo{volume}{D30}},
  \bibinfo{pages}{2295} (\bibinfo{year}{1984}).

\bibitem[{\citenamefont{Goodman and Witten}(1985)}]{Goodman:1984dc}
\bibinfo{author}{\bibfnamefont{M.~W.} \bibnamefont{Goodman}} \bibnamefont{and}
  \bibinfo{author}{\bibfnamefont{E.}~\bibnamefont{Witten}},
  \bibinfo{journal}{Phys.Rev.} \textbf{\bibinfo{volume}{D31}},
  \bibinfo{pages}{3059} (\bibinfo{year}{1985}).

\bibitem[{\citenamefont{Drukier et~al.}(1986)\citenamefont{Drukier, Freese, and
  Spergel}}]{Drukier:1986tm}
\bibinfo{author}{\bibfnamefont{A.}~\bibnamefont{Drukier}},
  \bibinfo{author}{\bibfnamefont{K.}~\bibnamefont{Freese}}, \bibnamefont{and}
  \bibinfo{author}{\bibfnamefont{D.}~\bibnamefont{Spergel}},
  \bibinfo{journal}{Phys.Rev.} \textbf{\bibinfo{volume}{D33}},
  \bibinfo{pages}{3495} (\bibinfo{year}{1986}).

\bibitem[{\citenamefont{Bernabei et~al.}(1999)}]{Bernabei:1998rv}
\bibinfo{author}{\bibfnamefont{R.}~\bibnamefont{Bernabei}} \bibnamefont{et~al.}
  (\bibinfo{collaboration}{DAMA}), \bibinfo{journal}{Il Nuovo Cimento A}
  \textbf{\bibinfo{volume}{112}}, \bibinfo{pages}{545} (\bibinfo{year}{1999}).

\bibitem[{\citenamefont{Bernabei et~al.}(2000)}]{Bernabei:2000qi}
\bibinfo{author}{\bibfnamefont{R.}~\bibnamefont{Bernabei}} \bibnamefont{et~al.}
  (\bibinfo{collaboration}{DAMA}), \bibinfo{journal}{Phys.Lett.}
  \textbf{\bibinfo{volume}{B480}}, \bibinfo{pages}{23} (\bibinfo{year}{2000}).

\bibitem[{\citenamefont{Bernabei et~al.}(2003)\citenamefont{Bernabei, Belli,
  Cappella, Cerulli, Montecchia et~al.}}]{Bernabei:2003za}
\bibinfo{author}{\bibfnamefont{R.}~\bibnamefont{Bernabei}},
  \bibinfo{author}{\bibfnamefont{P.}~\bibnamefont{Belli}},
  \bibinfo{author}{\bibfnamefont{F.}~\bibnamefont{Cappella}},
  \bibinfo{author}{\bibfnamefont{R.}~\bibnamefont{Cerulli}},
  \bibinfo{author}{\bibfnamefont{F.}~\bibnamefont{Montecchia}},
  \bibnamefont{et~al.}, \bibinfo{journal}{Riv.Nuovo Cim.}
  \textbf{\bibinfo{volume}{26N1}}, \bibinfo{pages}{1} (\bibinfo{year}{2003}),
  \eprint{astro-ph/0307403}.

\bibitem[{\citenamefont{Bernabei et~al.}(2008{\natexlab{a}})}]{Bernabei:2008yi}
\bibinfo{author}{\bibfnamefont{R.}~\bibnamefont{Bernabei}} \bibnamefont{et~al.}
  (\bibinfo{collaboration}{DAMA}), \bibinfo{journal}{Eur.Phys.J.}
  \textbf{\bibinfo{volume}{C56}}, \bibinfo{pages}{333}
  (\bibinfo{year}{2008}{\natexlab{a}}), \eprint{0804.2741}.

\bibitem[{\citenamefont{Bernabei et~al.}(2010)}]{Bernabei:2010mq}
\bibinfo{author}{\bibfnamefont{R.}~\bibnamefont{Bernabei}} \bibnamefont{et~al.}
  (\bibinfo{collaboration}{DAMA, LIBRA}), \bibinfo{journal}{Eur.Phys.J.}
  \textbf{\bibinfo{volume}{C67}}, \bibinfo{pages}{39} (\bibinfo{year}{2010}),
  \eprint{1002.1028}.

\bibitem[{\citenamefont{Bernabei
  et~al.}(2013{\natexlab{a}})\citenamefont{Bernabei, Belli, Cappella,
  Caracciolo, Castellano et~al.}}]{Bernabei:2013xsa}
\bibinfo{author}{\bibfnamefont{R.}~\bibnamefont{Bernabei}},
  \bibinfo{author}{\bibfnamefont{P.}~\bibnamefont{Belli}},
  \bibinfo{author}{\bibfnamefont{F.}~\bibnamefont{Cappella}},
  \bibinfo{author}{\bibfnamefont{V.}~\bibnamefont{Caracciolo}},
  \bibinfo{author}{\bibfnamefont{S.}~\bibnamefont{Castellano}},
  \bibnamefont{et~al.}, \bibinfo{journal}{Eur.Phys.J.}
  \textbf{\bibinfo{volume}{C73}}, \bibinfo{pages}{2648}
  (\bibinfo{year}{2013}{\natexlab{a}}), \eprint{1308.5109}.

\bibitem[{\citenamefont{Bernabei et~al.}(2012{\natexlab{a}})}]{Bernabei:2012wp}
\bibinfo{author}{\bibfnamefont{R.}~\bibnamefont{Bernabei}}
  \bibnamefont{et~al.}, \bibinfo{journal}{Eur. Phys. J.}
  \textbf{\bibinfo{volume}{C72}}, \bibinfo{pages}{2064}
  (\bibinfo{year}{2012}{\natexlab{a}}), \eprint{1202.4179}.

\bibitem[{\citenamefont{Shields et~al.}(2015)\citenamefont{Shields, Xu, and
  Calaprice}}]{Shields:2015wka}
\bibinfo{author}{\bibfnamefont{E.}~\bibnamefont{Shields}},
  \bibinfo{author}{\bibfnamefont{J.}~\bibnamefont{Xu}}, \bibnamefont{and}
  \bibinfo{author}{\bibfnamefont{F.}~\bibnamefont{Calaprice}},
  \bibinfo{journal}{Phys. Procedia} \textbf{\bibinfo{volume}{61}},
  \bibinfo{pages}{169} (\bibinfo{year}{2015}).

\bibitem[{\citenamefont{Amaré et~al.}(2016)}]{Amare:2016fmp}
\bibinfo{author}{\bibfnamefont{J.}~\bibnamefont{Amaré}} \bibnamefont{et~al.},
  in \emph{\bibinfo{booktitle}{{14th Marcel Grossmann Meeting on Recent
  Developments in Theoretical and Experimental General Relativity,
  Astrophysics, and Relativistic Field Theories (MG14) Rome, Italy, July 12-18,
  2015}}} (\bibinfo{year}{2016}), \eprint{1601.01184},
  \urlprefix\url{https://inspirehep.net/record/1413721/files/arXiv:1601.01184.pdf}.

\bibitem[{\citenamefont{Coarasa et~al.}(2017)}]{Coarasa:2017aol}
\bibinfo{author}{\bibfnamefont{I.}~\bibnamefont{Coarasa}} \bibnamefont{et~al.}
  (\bibinfo{year}{2017}), \eprint{1704.06861}.

\bibitem[{\citenamefont{Maruyama et~al.}(2017)}]{cosine}
\bibinfo{author}{\bibfnamefont{R.}~\bibnamefont{Maruyama}} \bibnamefont{et~al.}
  (\bibinfo{collaboration}{COSINE-100}), \emph{\bibinfo{title}{{COSINE-100 Dark
  Matter Experiment}}}, \bibinfo{howpublished}{\url{http://cosine.yale.edu}}
  (\bibinfo{year}{2017}), \bibinfo{note}{[Online; accessed 20-April-2017]}.

\bibitem[{\citenamefont{Barbosa~de Souza et~al.}(2017)}]{deSouza:2016fxg}
\bibinfo{author}{\bibfnamefont{E.}~\bibnamefont{Barbosa~de Souza}}
  \bibnamefont{et~al.} (\bibinfo{collaboration}{DM-Ice}),
  \bibinfo{journal}{Phys. Rev.} \textbf{\bibinfo{volume}{D95}},
  \bibinfo{pages}{032006} (\bibinfo{year}{2017}), \eprint{1602.05939}.

\bibitem[{\citenamefont{Kim et~al.}(2012)}]{Kim:2012rza}
\bibinfo{author}{\bibfnamefont{S.~C.} \bibnamefont{Kim}} \bibnamefont{et~al.},
  \bibinfo{journal}{Phys. Rev. Lett.} \textbf{\bibinfo{volume}{108}},
  \bibinfo{pages}{181301} (\bibinfo{year}{2012}), \eprint{1204.2646}.

\bibitem[{\citenamefont{Bernabei et~al.}(2013{\natexlab{b}})}]{Bernabei:2013qx}
\bibinfo{author}{\bibfnamefont{R.}~\bibnamefont{Bernabei}}
  \bibnamefont{et~al.}, in \emph{\bibinfo{booktitle}{{Proceedings, 15th
  Workshop on What Comes Beyond the Standard Models?}}}
  (\bibinfo{year}{2013}{\natexlab{b}}), pp. \bibinfo{pages}{1--9},
  \bibinfo{note}{[,1(2012)]}, \eprint{1301.6243},
  \urlprefix\url{http://inspirehep.net/record/1216452/files/arXiv:1301.6243.pdf}.

\bibitem[{\citenamefont{Kelso et~al.}(2013)\citenamefont{Kelso, Sandick, and
  Savage}}]{Kelso:2013gda}
\bibinfo{author}{\bibfnamefont{C.}~\bibnamefont{Kelso}},
  \bibinfo{author}{\bibfnamefont{P.}~\bibnamefont{Sandick}}, \bibnamefont{and}
  \bibinfo{author}{\bibfnamefont{C.}~\bibnamefont{Savage}},
  \bibinfo{journal}{JCAP} \textbf{\bibinfo{volume}{1309}}, \bibinfo{pages}{022}
  (\bibinfo{year}{2013}), \eprint{1306.1858}.

\bibitem[{\citenamefont{Freese et~al.}(1988)\citenamefont{Freese, Frieman, and
  Gould}}]{Freese:1987wu}
\bibinfo{author}{\bibfnamefont{K.}~\bibnamefont{Freese}},
  \bibinfo{author}{\bibfnamefont{J.~A.} \bibnamefont{Frieman}},
  \bibnamefont{and} \bibinfo{author}{\bibfnamefont{A.}~\bibnamefont{Gould}},
  \bibinfo{journal}{Phys.Rev.} \textbf{\bibinfo{volume}{D37}},
  \bibinfo{pages}{3388} (\bibinfo{year}{1988}).

\bibitem[{\citenamefont{Freese et~al.}(2013)\citenamefont{Freese, Lisanti, and
  Savage}}]{Freese:2012xd}
\bibinfo{author}{\bibfnamefont{K.}~\bibnamefont{Freese}},
  \bibinfo{author}{\bibfnamefont{M.}~\bibnamefont{Lisanti}}, \bibnamefont{and}
  \bibinfo{author}{\bibfnamefont{C.}~\bibnamefont{Savage}},
  \bibinfo{journal}{Rev.Mod.Phys.} \textbf{\bibinfo{volume}{85}},
  \bibinfo{pages}{1561} (\bibinfo{year}{2013}), \eprint{1209.3339}.

\bibitem[{\citenamefont{Bernabei et~al.}(2008{\natexlab{b}})}]{Bernabei:2008yh}
\bibinfo{author}{\bibfnamefont{R.}~\bibnamefont{Bernabei}} \bibnamefont{et~al.}
  (\bibinfo{collaboration}{DAMA}), \bibinfo{journal}{Nucl.Instrum.Meth.}
  \textbf{\bibinfo{volume}{A592}}, \bibinfo{pages}{297}
  (\bibinfo{year}{2008}{\natexlab{b}}), \eprint{0804.2738}.

\bibitem[{\citenamefont{Bernabei
  et~al.}(2012{\natexlab{b}})\citenamefont{Bernabei, Belli, Bussolotti,
  Cappella, Caracciolo, Casalboni, Cerulli, Dai, d'Angelo, Marco
  et~al.}}]{1748-0221-7-03-P03009}
\bibinfo{author}{\bibfnamefont{R.}~\bibnamefont{Bernabei}},
  \bibinfo{author}{\bibfnamefont{P.}~\bibnamefont{Belli}},
  \bibinfo{author}{\bibfnamefont{A.}~\bibnamefont{Bussolotti}},
  \bibinfo{author}{\bibfnamefont{F.}~\bibnamefont{Cappella}},
  \bibinfo{author}{\bibfnamefont{V.}~\bibnamefont{Caracciolo}},
  \bibinfo{author}{\bibfnamefont{M.}~\bibnamefont{Casalboni}},
  \bibinfo{author}{\bibfnamefont{R.}~\bibnamefont{Cerulli}},
  \bibinfo{author}{\bibfnamefont{C.~J.} \bibnamefont{Dai}},
  \bibinfo{author}{\bibfnamefont{A.}~\bibnamefont{d'Angelo}},
  \bibinfo{author}{\bibfnamefont{A.~D.} \bibnamefont{Marco}},
  \bibnamefont{et~al.}, \bibinfo{journal}{Journal of Instrumentation}
  \textbf{\bibinfo{volume}{7}}, \bibinfo{pages}{P03009}
  (\bibinfo{year}{2012}{\natexlab{b}}),
  \urlprefix\url{http://stacks.iop.org/1748-0221/7/i=03/a=P03009}.

\end{thebibliography}

\end{document}